\documentclass[aps,twocolumn,groupedaddress,superscriptaddress,showpacs]{revtex4}
\usepackage{graphicx}
\usepackage{colortbl}
\makeatletter
\def\btt#1{\texttt{\@backslashchar#1}}%
\DeclareRobustCommand\bblash{\btt{\@backslashchar}}%
\makeatother

\begin{document}

%\baselineskip=2\normalbaselineskip
%���P�s�󂫂ɂ����ꍇ

\title{ Size effects on supercooling phenomena in strongly correlated electron systems: IrTe$_2$ and $\theta$-(BEDT-TTF)$_2$RbZn(SCN)$_4$}

\author{H. Oike}
\email{hiroshi.oike@riken.jp}
\affiliation{RIKEN Center for Emergent Matter Science (CEMS), Wako 351-0198, Japan}

\author{M. Suda}
\affiliation{Research Center for Integrative Molecular System (CIMoS), Institute for Molecular Science, Okazaki 444-8585, Japan}

\author{M. Kamitani}
\affiliation{RIKEN Center for Emergent Matter Science (CEMS), Wako 351-0198, Japan}

\author{A. Ueda}
\affiliation{The Institute for Solid State Physics, The University of Tokyo, Kashiwa 277-8581, Japan}

\author{H. Mori}
\affiliation{The Institute for Solid State Physics, The University of Tokyo, Kashiwa 277-8581, Japan}

\author{Y. Tokura}
\affiliation{RIKEN Center for Emergent Matter Science (CEMS), Wako 351-0198, Japan}
\affiliation{Department of Applied Physics, The University of Tokyo, Tokyo 113-8656, Japan}

\author{H. M. Yamamoto}
\affiliation{Research Center for Integrative Molecular System (CIMoS), Institute for Molecular Science, Okazaki 444-8585, Japan}

\author{F. Kagawa}
\email{fumitaka.kagawa@riken.jp}
\affiliation{RIKEN Center for Emergent Matter Science (CEMS), Wako 351-0198, Japan}
\affiliation{Department of Applied Physics, The University of Tokyo, Tokyo 113-8656, Japan}

\date{\today}
\begin{abstract}
We report that the sample miniaturization of first-order-phase-transition bulk systems causes a greater degree of supercooling. From a theoretical perspective, the size effects can be rationalized by considering two mechanisms: (i) the nucleation is a rare and stochastic event, and thus, its rate is correlated with the volume and/or surface area of a given sample; (ii) when the sample size decreases, the dominant heterogeneous nucleation sites that play a primary role for relatively large samples are annealed out. We experimentally verified the size effects on the supercooling phenomena for two different types of strongly correlated electron systems: the transition-metal dichalcogenide IrTe$_{2}$ and the organic conductor $\theta$-(BEDT-TTF)$_2$RbZn(SCN)$_4$. The origin of the size effects considered in this study does not depend on microscopic details of the material; therefore, they may often be involved in the first-order-transition behavior of small-volume specimens.
\end{abstract}

\pacs{xxxx}

\maketitle

%***********************************************************************
\section{Introduction}

	Most first-order phase transitions are initiated by nucleation, which is a critical rare event that changes the manner in which a system subsequently develops over time \cite{Frenkel, GasserScience}. In general, the nucleation rate strongly depends on the temperature $T$, and its temperature evolution is often non-monotonic, as shown in Fig.~1(a). This non-monotonic behavior has various consequences for the macroscopic progression of a first-order phase transition that occurs under continuous cooling at a given rate $r (> 0)$, as schematically shown in Fig.~1(b). First, the first-order transitions do not occur until supercooling progresses to a certain extent. This universal propensity is traced to the fact that the nucleation rate is minuscule at temperatures immediately below a thermo-equilibrium first-order transition temperature $T_{\rm c}$, which is defined as a temperature at which the free energies of the high- and low-temperature phases become equal. Second, a higher cooling rate results in a greater degree of supercooling because the time spent at each temperature point becomes more limited. Thus, the experimental transition temperature on cooling $T_{\rm c}^{\rm cool}$ is cooling-rate-dependent. Third, even higher cooling rates eventually cause a persistent supercooling to the lowest temperature because the nucleation rate again becomes minuscule at sufficiently low temperatures. The most extensively studied realization of this scenario is classical liquids [Fig.~1(c)] \cite{Uhlmann, AngellScience, DebenedettiNaure}: although all pure and simple liquids tend to crystalize at a cooling-rate-dependent temperature, this first-order crystallization transition can be kinetically avoided under sufficiently rapid cooling. As a result, a non-ergodic structural glass forms below a glass transition temperature $T_{\rm g}^\ast$, which also depends on the cooling rate.

%%%%%%%%%%%%%%%%%%%Fig1
\begin{figure}
\includegraphics[width=5.8cm]{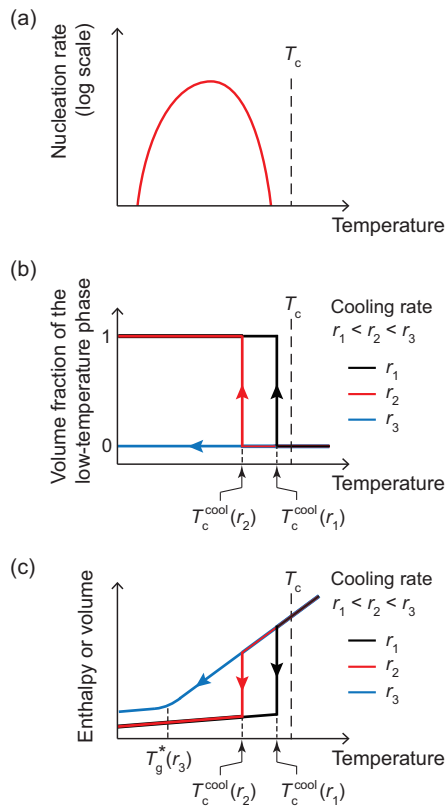}
\caption{\label{Fig1}
(Color online) (a) Schematic of the temperature dependence of the nucleation rate. (b) Corresponding temperature dependence of the volume fraction of the low-temperature phase under various cooling rates $r (> 0)$. As the cooling rate increases ($r_1 < r_2 < r_3$), the first-order transition occurs at a lower temperature. At the highest cooling rate $r_3$, the first-order transition is kinetically avoided to the lowest temperature. (c) Corresponding enthalpy- (or volume-) temperature profiles of classical liquids. }

\end{figure}
%%%%%%%%%%%%%%%%%%%Fig1

Sample miniaturization has been recognized to also be helpful in obtaining deeply supercooled liquids \cite{Turnbull} or metastable crystalline solids \cite{BrusJACS, ScienceChen}. Overall, the mechanism of this size effect is two-fold. First, when the sample size decreases, the total number of involved nucleation sites decreases accordingly, which results in a lower chance for nucleation. Second, in a sufficiently small sample, the dominant heterogeneous nucleation sites that play a primary role for relatively large samples are annealed out. Both mechanisms inhibit nucleation and thus facilitate a prolonged lifetime for metastable solids.
	
The idea of rapid cooling was recently extended and applied to first-order transitions in electronic degrees of freedom in quantum materials (for a review, see ref.~\cite{QuenchReview}). With the use of optical/electronic pulses, an unconventionally high cooling rate such as 10$^2$$-$10$^3$ K/s is easily achieved, which enables certain quantum materials to kinetically avoid a first-order phase transition in the charge \cite{KagawaNatPhys, OikePRB} or spin \cite{HoMo6S8JMMM, HoMo6S8SSC, HoMo6S8Physica, OikeNatPhys, MotoyaJPSJ, NakajimaSciAdv, KagawaNatCommun} degrees of freedom. As a result, a quenched electronic/magnetic state that differs from the ground state is realized at low temperatures; such quenched states can also be used as a non-volatile state variable in the design of phase-change memory functions \cite{OikePRB, OikeNatPhys}. However, the possible role of the sample size in supercooling phenomena has been examined only in limited quantum materials, such as VO$_2$, in which a greater degree of supercooling is observed for a smaller sample \cite{LopezPRB, VO2}. In this Article, we investigate the size effect for two distinct quantum materials: the transition-metal dichalcogenide IrTe$_{2}$ and the organic conductor $\theta$-(BEDT-TTF)$_2$RbZn(SCN)$_4$, where BEDT-TTF denotes bis(ethylenedithio)tetrathiafulvalene.

%***********************************************************************
\section{Numerical simulations}

Before presenting the experimental results, we present an overview of the correlations among supercooling phenomena, which include the kinetic avoidance of first-order transitions, rapid cooling, and/or sample miniaturization.

\subsection{Phenomenological description of the nucleation rate in a given mass}

The nucleation rate in a given mass $I(T)$ [s$^{-1}$] is generally dictated by the sum of the homogeneous and heterogeneous nucleation rates in the bulk and the nucleation rates at the surfaces. Heterogeneous nucleation indicates nucleation at defects, such as impurity atoms, vacancies, and dislocations. For simplicity, we represent the nucleation rate at a certain type of defect with a single value $I_{{\rm defect}, j}(T)$ [s$^{-1}$], where $j$ is an index that specifies the type of defect. Thus, the total heterogeneous nucleation rate in the bulk is approximately given by $\sum_{j}I_{{\rm defect},j}(T)N_{{\rm defect},j}$, where $N_{{\rm defect},j}$ is the total number of $j$ defects contained in the sample and should be proportional to the sample volume $V$, as long as defects are uniformly distributed in the sample. Therefore, the heterogeneous nucleation rate can be normalized with respect to $V$, and the normalized value is called the heterogeneous nucleation rate density and defined as $i_{{\rm v},j}^{\rm hetero}(T) \equiv I_{{\rm defect},j}(T)\rho_{{\rm defect},j}$ [s$^{-1}$~m$^{-3}$], where $\rho_{{\rm defect},j}$ is the density of $j$ defects, $N_{{\rm defect},j}/V$. Similarly, one can also consider the homogeneous nucleation rate density $i_{\rm v}^{\rm homo}(T)$ [s$^{-1}$~m$^{-3}$]. The nucleation rates at the sample surface is proportional to the surface area $S$; thus, the normalized surface nucleation rate is defined as $i_{\rm s}(T)$ [s$^{-1}$~m$^{-2}$]. 

Thus, for example, the total nucleation rate $I(T)$ in a rectangular-parallelepiped-shaped sample with x-, y-, and z-planes is approximated as:
\begin{equation}
I(T)=V\{i_{\rm v}^{\rm homo}(T) + \sum_{j}i_{{\rm v}, j}^{\rm hetero}(T)\} + \sum_{k=x,y,z}S_{k}i_{{\rm s}, k}(T),
\end{equation}
where $S_{k}$ and $i_{{\rm s}, k}$ represent the surface area and the normalized surface nucleation rate, respectively, for the $k$-plane ($k$ = x, y, z) \cite{Eq1}. Which term is dominant depends on the system details, and the resolution of this issue is beyond the scope of this paper. Nevertheless, Eq.~(1) explicitly shows that nucleation events become less frequent as the sample size is decreased in terms of both $V$ and $S$, a natural consequence of the fact that nucleation is a rare event in a given system. The purpose of the following section is to show that, in regards to the supercooling phenomena, sample miniaturization is qualitatively similar to the application of rapid cooling.

\subsection{First nucleation event during continuous cooling}

To simulate a first nucleation event that occurs during continuous cooling at a given rate $r (>0)$, we set $t=0$ as the moment when the simulation temperature passes through the thermodynamic $T_{\rm c}$. The probability $p(t)dt$ that the first nucleation event occurs at a time interval between $t$ and $t+dt$ is extracted from the Poisson distribution:
\begin{equation}
p(t)dt = I(T(t))dt \exp \left[-\int_0^{t}I(T(t'))dt' \right],
\end{equation}
where exp[$\cdots$] is the probability that no nucleation event occurs until the considered time interval is reached. Hence, the probability $P(T)dT$ that the first nucleation event occurs at a temperature interval between $T$ and $T-dT$ is
\begin{equation}
P(T)dT = (I(T)/r)dT \exp \left[-\int_{T}^{T_{\rm c}} (I(T')/r)dT' \right].
\end{equation}
Note that the nucleation rate $I(T)$ and cooling rate $r$ in Eq.~(3) always appear as a pair in the form $I(T)/r$, which, regarding the probability of the first nucleation, explicitly shows the equivalence between the suppression of $I(T)$ and the application of a higher cooling rate. As described in Eq.~(1), sample miniaturization in terms of both $V$ and $S$ is a facile method to decrease the $I(T)$ of a specimen.

\subsection{Exemplary studies}

%%%%%%%%%%%%%%%%%%%Fig2
\begin{figure}
\includegraphics[width=5.1cm]{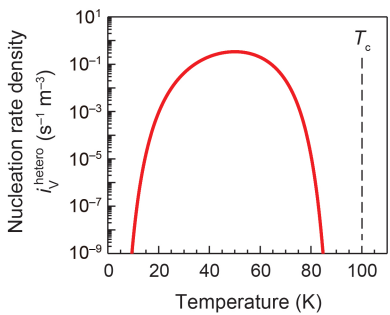}
\caption{\label{Fig2}
(Color online) Modeled temperature dependence of the nucleation rate density $I_{\rm v}(T)$ for the simulation. The linear-scale representation is shown in Supplementary Fig.~S1. For the details of the functional form, see ref.~\cite{Simulation}. 
}

\end{figure}
%%%%%%%%%%%%%%%%%%%Fig2

%%%%%%%%%%%%%%%%%%%Fig3
\begin{figure}
\includegraphics[width=7.8cm]{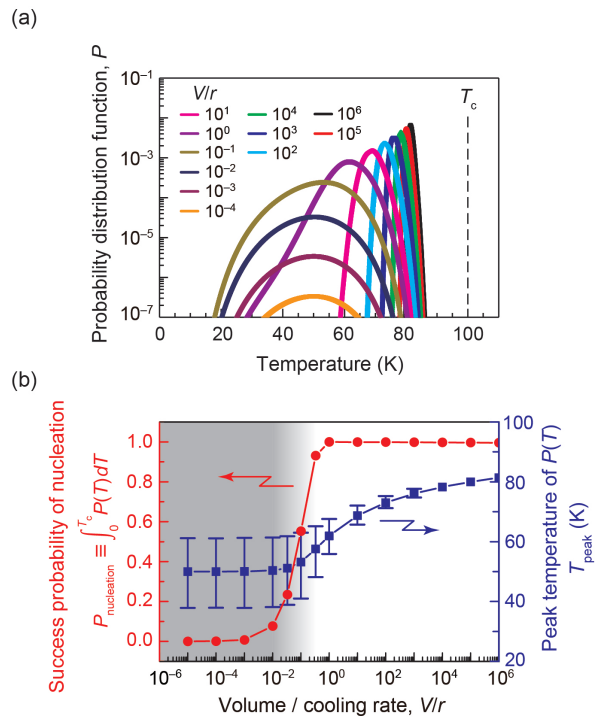}
\caption{\label{Fig3}
(Color online) (a) Calculated probability distribution function $P(T)$ regarding the occurrence of the first nucleation event. The calculation was performed according to Eq.~(4). The linear-scale representation is shown in Supplementary Fig.~S2. (b) $V/r$ dependence of the success probability of nucleation during cooling to zero temperature (the left axis) and the peak temperature of $P(T)$ (the right axis). The error bars correspond to the full-width of half-maximum of $P(T)$ in Fig.~3(a). The unit of $V/r$ is [K$^{-1}$~s~m$^{3}$].
}

\end{figure}
%%%%%%%%%%%%%%%%%%%Fig3

To gain further insights into the impact of sample miniaturization on supercooling phenomena, we address the simplest case where nucleation occurs exclusively at a certain type of defect in the bulk, namely, $I(T) \approx Vi_{\rm v}^{\rm hetero}(T)$. Thus, Eq.~(3) is rewritten as:
\begin{equation}
P(T)dT = i_{\rm v}^{\rm hetero}(T)(V/r)dT \exp \left[-(V/r) \int_{T}^{T_{\rm c}} i_{\rm v}^{\rm hetero}(T') dT' \right].
\end{equation}
In this study, we modeled $i_{\rm v}^{\rm hetero}(T)$, as shown in Fig.~2 ($T_{\rm c}=100$ K; for more details, see ref.~\cite{Simulation, hosoku}), and calculated $P(T)$ for various $V/r$, as shown in Fig.~3(a). The peak temperature $T_{\rm peak}$ in each $P(T)$ curve is plotted as a function of $V/r$ in the right axis of Fig.~3(b), where the error bars represent the full-width of half-maximum. To determine whether nucleation events can statistically be expected during cooling to zero temperature, we also calculated $P_{\rm nucleation} \equiv \int_{0}^{T_{\rm c}} P(T)dT$, as shown in the left axis of Fig.~3(b).

As an exemplary approach, we consider the case where the growth speed of the phase front of the post-critical nucleus is sufficiently fast such that a single nucleation event immediately completes a first-order phase transition in the whole sample volume. In this case, the transition is manifested as an abrupt jump, as schematically shown in Fig.~1(b), and supercooling phenomena are well represented by the kinetics of first nucleation, which is summarized in Fig.~3(b). Overall, the transition behavior that varies with $V/r$ is categorized into two regimes: ``the slow-cooling regime'', in which the first-order transition (abruptly) occurs after certain supercooling, and ``the quenched regime'', in which the transition is kinetically avoided to the lowest temperature \cite{QuenchReview}. In Fig.~3(b), the slow-cooling regime encompasses a relatively large $V/r$ (e.g., $V/r > 0.33$), which corresponds to a slow-cooling experiment and/or a large sample. In this $V/r$ range, $P_{\rm nucleation}$ is greater than 0.9; thus, the transition almost invariably occurs during cooling at a temperature near $T_{\rm peak}$. In the large $V/r$ limit, $T_{\rm peak}$ asymptotically approaches the thermodynamic $T_{\rm c} (=100$ K). When $V/r$ decreases from the large limit, $T_{\rm peak}$ is increasingly lowered [Fig.~3(b)] and eventually reaches the peak temperature of $i_{\rm v}^{\rm hetero}(T)$ ($\approx$50 K). The quenched regime then sets in (e.g., $V/r < 10^{-2}$), where the transition is statistically not expected because of the low values of $P_{\rm nucleation}$ (less than 0.1). When the surface nucleation is dominant (i.e., $I(T) \approx Si_{\rm s}(T)$), $V$ in the above discussion should be substituted with $S$, but such a modification does not affect the qualitative conclusion above, namely, the slow-cooling and quenched regimes emerge as a function of $S/r$.

%%%%%%%%%%%%%%%%%%%Fig4
\begin{figure}
\includegraphics[width=7.0cm]{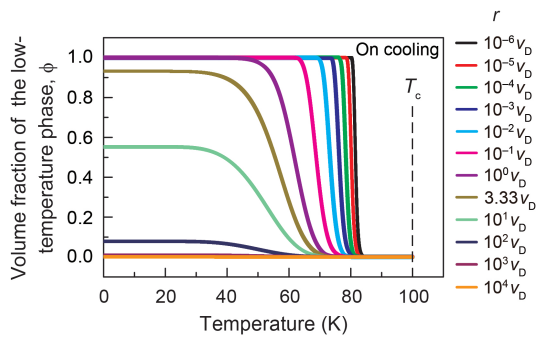}
\caption{\label{Fig4}
(Color online) Temperature dependence of the calculated volume fraction of the low-temperature phase, which consists of multiple domains with a volume of $v_{\rm D}$. The calculation is performed according to Eq.~(6), and the considered cooling rate is displayed with respect to $v_{\rm D}$ multiplied by a certain coefficient, the unit of which is [K~s$^{-1}$~m$^{-3}$].
}
\end{figure}
%%%%%%%%%%%%%%%%%%%Fig4

However, one should note that the size effect in real systems would not be as simple as that discussed. Below, we show two extreme cases where the sample miniaturization does not affect supercooling at all. The first example is when the growth speed is zero. In this case, at a given $T$, the ratio of nucleated sites to the total potential nucleation sites $a(T)$ is given as
\begin{equation}
a(T) = 1- \exp \left[-\int_{T}^{T_{\rm c}} I_{\rm defect}(T')/r dT' \right].
\end{equation}
Obviously, Eq.~(5) is $V$ independent, indicating that, in the limit of zero growth speed, no impact of the sample miniaturization is expected. Nevertheless, $a(T)$ depends on $r$ (for details, see Supplementary Fig.~S3).

The second example is when, although the growth speed is sufficiently fast, the continual growth of the post-critical nucleus is stopped at certain pinning sites of extrinsic origin. In this case, each nucleation event is accompanied by the phase transition of only a certain volume, and thus, the low-temperature state consists of multiple domains. Hence, the temperature evolution of the low-temperature-phase volume fraction $\phi$$(T)$ is given as
\begin{equation}
\phi(T) = 1- \exp \left[-(v_{\rm D}/r) \int_{T}^{T_{\rm c}} i_{\rm v}^{\rm hetero}(T') dT' \right],
\end{equation}
where $v_{\rm D}$ is the typical domain size and $\rho_{\rm defect}^{-1} \ll v_{\rm D} \ll V$ is assumed for simplicity. Note that, in this regime, $\phi(T)$ is independent of the macroscopic volume $V$. The temperature profile of $\phi(T)$ is continuous (Fig.~4) because numerous nucleation events are involved in the phase transition process; therefore, the degree of supercooling can be defined only approximately (for more details, see Supplementary Fig.~S4). Nevertheless, it is graphically clear that a higher cooling rate tends to shift the characteristic transition-temperature range toward lower temperatures. Thus, similar to the case of Fig.~3(b), the slow-cooling and quenched regimes emerge, with the cooling rate as the only control parameter (Supplementary Fig.~S4). When $V < v_{\rm D}$ (i.e., in the monodomain regime), the size effect should appear, as discussed in Fig.~3(b).

Up to this point, we have shown that although the size effect appears when a single nucleation event is sufficient to complete the phase transition, the sample size becomes irrelevant when numerous nucleation events are involved even in the slow-cooling limit and, thus, the impact of the first nucleation is minuscule. Real systems are mostly in an intermediate state between these extreme conditions, namely, the nucleation events involved are neither single nor numerous, and the growth speed is neither zero nor infinite. Thus, the sample miniaturization in terms of both $V$ and $S$ is expected to more or less affect the degree of supercooling, but its dependence would be less pronounced than expected from Eqs.~(1) and (3). The origin of the size effect is ultimately traced to the fact that the initiating process of the first-order phase transition, nucleation, is a rare and stochastic process.

\subsection{Crossover regarding the nucleation mechanism}

As described in Eq.~(1), the dominant nucleation mechanism can vary with sample size and shape. Although the bulk nucleation should be dominant in a sufficiently large sample (i.e., $I(T) \sim V$), it would be superseded by surface nucleation in a sufficiently small sample (i.e., $I(T) \sim S \sim V^{2/3}$) because of an enhanced surface-to-volume ratio. When such a bulk-to-surface crossover occurs upon sample miniaturization, the surface area would be a more appropriate parameter to consider.

So far, we have implicitly supposed that Eq.~(1) holds upon sample miniaturization. However, this is not always the case: when $V < \rho_{\rm defect}^{-1}$, the small sample does not contain the defects that play a primary role in nucleation for a large sample. In such a case, the second dominant term for a large sample becomes the primary term for a sufficiently small sample. In this crossover regime associated with heterogeneous nucleation, $I(T)/r$ in Eq.~(3) may exhibit a more pronounced change under $V$ variation than $r$ variation when either $V$ or $r$ is changed, for instance, by one order of magnitude. This situation appears to be relevant to the case of IrTe$_2$ (see below).

%***********************************************************************
\section{Experimental results}

To experimentally test the discussed possible size effects, it is imperative to track how the degree of supercooling $T_{\rm c}^{\rm cool}$ systematically varies with the sample size and/or cooling rate. In this study, we targeted two different materials: the transition-metal dichalcogenide IrTe$_{2}$ and the organic conductor $\theta$-(BEDT-TTF)$_2$RbZn(SCN)$_4$. Both materials exhibit a first-order transition accompanied by a clear resistivity change; thus, standard resistivity measurements enable us to address the change of $T_{\rm c}^{\rm cool}$, which is defined as the onset temperature of the transition. For the details of the sample preparation, see the Supplementary Material section.

\subsection{Transition-metal dichalcogenide, IrTe$_{2}$}

%%%%%%%%%%%%%%%%%%%Fig5
\begin{figure}
\includegraphics[width=5.6cm]{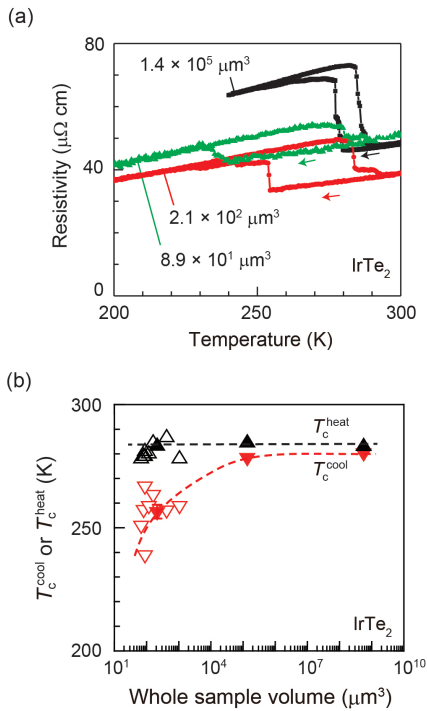}
\caption{\label{Fig5}
(Color online) (a) Temperature dependence of the resistivity at a fixed sweep rate, 3 K/min, for three IrTe$_2$ crystals with different volumes. (b) $T_{\rm c}^{\rm cool}$ and $T_{\rm c}^{\rm heat}$ variations with the whole sample volume. The red and black symbols represent $T_{\rm c}^{\rm cool}$ and $T_{\rm c}^{\rm heat}$, respectively. $T_{\rm c}^{\rm cool}$ and $T_{\rm c}^{\rm heat}$ are defined as the onset temperature of the transition (for details, see Supplementary Fig. S5). For select samples (closed symbols), we examined the reproducibility of the transition temperature by repeating certain experiments; the observed variations in $T_{\rm c}^{\rm cool}$ or $T_{\rm c}^{\rm heat}$ are represented by error bars. The broken lines are drawn as a guide for the eyes. For the graph plotted against the sample surface area, see Supplementary Fig. S6.
}

\end{figure}
%%%%%%%%%%%%%%%%%%%Fig5

IrTe$_2$ has a CdI$_2$-type layered structure that consists of stacked layers of IrTe$_6$ octahedra. Upon cooling, this material exhibits two successive first-order phase transitions at $\approx$280 K and $\approx$180 K, which are associated with the coupling between the lattice and the charge \cite{EomPRL, KoNatCommun, ChenPRB}. In this study, we examined the sample-miniaturization effect on the transition at $\approx$280 K, where the unit cell changes from the original $1\times$$1\times$1 structure into a $5\times$$1\times$5 structure, which has a five-fold greater periodicity  in the $a$ and $c$ axes \cite{YangPRL, MazumdarPRB}.

Figure 5(a) shows the temperature dependence of the resistivity measured at a fixed rate of 3 K/min for three selected IrTe$_{2}$ samples with different volumes ($1.4\times$10$^5$, $2.1\times$10$^2$, and $1.9\times$10$^1$ $\mu$m$^3$, the surface areas of which are $4.2\times$10$^4$, $8.3\times$10$^2$, and $6.2\times$10$^2$ $\mu$m$^2$, respectively); $T_{\rm c}^{\rm cool}$ for all of the examined samples is summarized in terms of the volume in Fig.~5(b) (for the graph plotted against the sample surface area, see Supplementary Fig.~S6). Note that $T_{\rm c}^{\rm cool}$ systematically shifts toward low temperatures when the volume decreases. Although $T_{\rm c}^{\rm cool}$ varies from one cooling cycle to the next (see Supplementary Fig.~S7) and should therefore be discussed with a certain error bar, the largeness of the degree of supercooling is well above this uncertainty [Fig.~5(b)]. Additionally, the cooling-rate dependence was not clearly observed beyond the error bar for all of the examined samples (Supplementary Fig.~S7). Thus, in the examined samples and cooling-rate range, the $V$ dependence (or $S$ dependence) of supercooling is more pronounced than the $r$ dependence. This behavior is explained by supposing that the dominant term in Eq.~(1) changes as the sample size decreases: the crossover regarding the dominant nucleation mechanism may be hetero-to-hetero, hetero-to-homo, or bulk-to-surface. We also note a salient size effect on $T_{\rm c}^{\rm cool}$ in the other transition-metal dichalcogenide 1$T$-TaS$_{2}$ \cite{YoshidaSciRep, YoshidaSciAdv}, regarding a first-order transition between ``the nearly commensurate charge density wave (NCCDW)'' and ``the commensurate charge density wave (CCDW)''; however, the size effect on the nucleation probability has not been discussed.

Although our main focus in this study is supercooling phenomena that accompany the sample miniaturization, we also plotted $T_{\rm c}^{\rm heat}$ in Fig.~5(b) and found that in contrast to $T_{\rm c}^{\rm cool}$, $T_{\rm c}^{\rm heat}$ is much less sensitive to the sample volume (similar to 1$T$-TaS$_{2}$ \cite{YoshidaSciRep, YoshidaSciAdv}). The nearly constant $T_{\rm c}^{\rm heat}$ presumably indicates that the thermodynamic $T_{\rm c}$ is not substantially affected by the sample miniaturization in the present range. At a minimum, in ordinary liquids, a ``crystallization temperature'' ($\equiv T_{\rm c}^{\rm cool}$) can widely vary with the experimental cooling rates and/or boundary conditions such as a container, whereas the ``melting temperature'' ($\equiv T_{\rm c}^{\rm heat}$) is notably insensitive to them \cite{textbook}.

\subsection{Organic conductor, $\theta$-(BEDT-TTF)$_2$RbZn(SCN)$_4$}

%%%%%%%%%%%%%%%%%%%Fig6
\begin{figure}
\includegraphics[width=8.6cm]{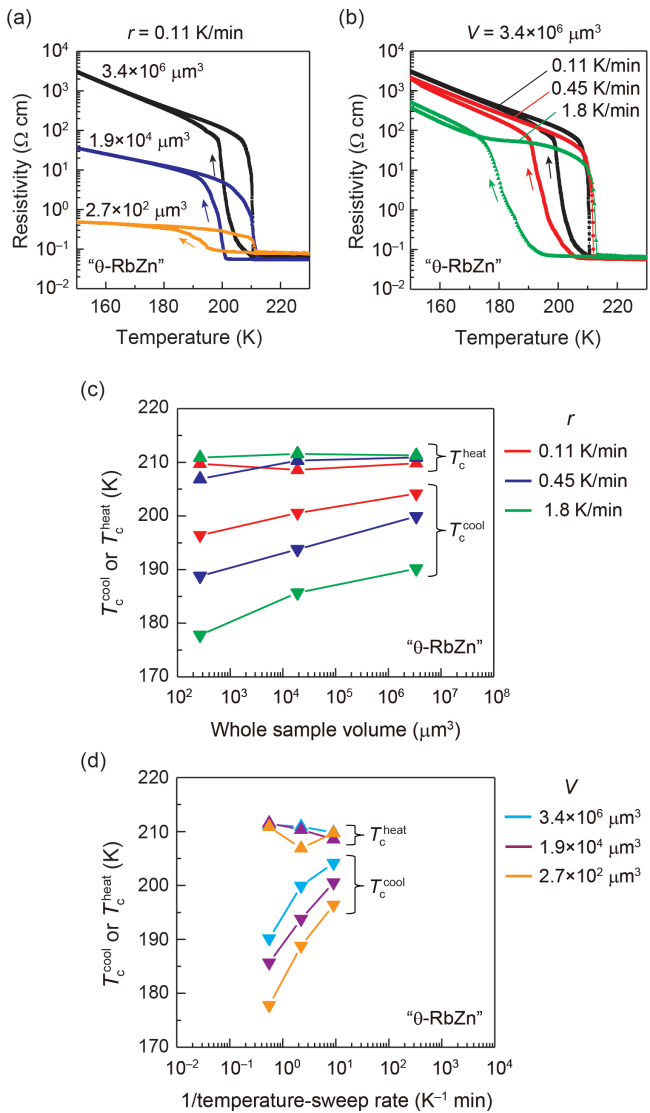}
\caption{\label{Fig6}
(Color online) (a) Temperature dependence of the resistivity at a fixed sweep rate of 0.11 K/min for three $\theta$-(BEDT-TTF)$_2$RbZn(SCN)$_4$ crystals with different volumes. (b) Temperature dependence of the resistivity measured at different sweeping rates for the sample with $V=3.4\times10^6$ $\mu$m$^3$. (c,d) $T_{\rm c}^{\rm cool}$ and $T_{\rm c}^{\rm heat}$ variations with the (c) whole sample volume and (d) cooling rate. The temperature-resistivity profile is highly reproducible upon repeated thermal cycling (Supplementary Fig.~S13) at least for the most supercooled sample ($V=2.7\times10^2$ $\mu$m$^3$); thus, the error bars of $T_{\rm c}^{\rm cool}$ and $T_{\rm c}^{\rm heat}$ are within the symbol size. For the graph plotted against the surface area, see Supplementary Fig.~S9.
}

\end{figure}
%%%%%%%%%%%%%%%%%%%Fig6

We performed similar experiments for $\theta$-(BEDT-TTF)$_2$RbZn(SCN)$_4$ (abbreviated as $\theta$-RbZn). The crystal structure consists of alternating layers of conducting BEDT-TTF molecules and insulating RbZn(SCN)$_4$. Upon cooling, this material exhibits a first-order phase transition from a semiconducting charge-itinerant state to an insulating charge-ordered state at 190--200 K \cite{KagawaNatPhys, HatsumiPRB, WatanabeJPSJ, MiyagawaPRB, YamamotoPRB}. Because of the charge ordering, the structure of the conducting $a$-$c$ plane changes to a 1$\times$2 structure (i.e., a two-fold greater periodicity in the $c$ axis), and a horizontal charge stripe with a charge disproportionation ratio of $0.15:0.85$ is formed \cite{WatanabeJPSJ, MiyagawaPRB, YamamotoPRB}. This charge ordering is accompanied by a large resistivity increase of several orders of magnitude.

Figure 6(a) shows the temperature dependence of the resistivity measured at a fixed sweep rate of 0.11 K/min for three $\theta$-RbZn samples with different volumes ($3.4\times$10$^6$, $1.9\times$10$^4$, and $2.7\times$10$^2$ $\mu$m$^3$, the surface areas of which are $3.6\times$10$^5$, $4.0\times$10$^4$, and $1.0\times$10$^3$ $\mu$m$^2$, respectively). As in the case of IrTe$_2$, $T_{\rm c}^{\rm cool}$ systematically shifts toward low temperatures when the volume decreases, whereas $T_{\rm c}^{\rm heat}$ is nearly volume-independent. The temperature-sweep-rate-dependent variations of the transition behavior are displayed for the largest sample in Fig.~6(b). A higher cooling rate reasonably results in a larger degree of supercooling, whereas $T_{\rm c}^{\rm heat}$ is insensitive to the heating rate. The temperature-sweep-rate dependences for the other two samples are shown in Supplementary Fig.~S8.

Figure 6(c) summarizes the volume dependences of $T_{\rm c}^{\rm cool}$ and $T_{\rm c}^{\rm heat}$ measured at different temperature-sweep rates (for the graph plotted against the surface area, see Supplementary Fig.~S9). The same data sets are also displayed as a function of the cooling rate in Fig.~6(d). $T_{\rm c}^{\rm cool}$ and $T_{\rm c}^{\rm heat}$ are defined as the onset temperatures of the transitions seen in the logarithmic scale (for details, see Supplementary Fig.~S10. If $T_{\rm c}^{\rm cool}$ and $T_{\rm c}^{\rm heat}$ are defined in the linear-scale conductivity plot, the values of $T_{\rm c}^{\rm cool}$ and $T_{\rm c}^{\rm heat}$ change slightly. Nevertheless, the global tendency of Fig.~6(c) is affected only weakly; see Supplementary Fig.~S11 and S12). A comparison of Figs.~6(c) and (d) shows that, in contrast to the case of IrTe$_2$, $T_{\rm c}^{\rm cool}$ more clearly depends on $r$ than on $V$ (or $S$) in $\theta$-RbZn. This tendency implies that multiple nucleation events are involved in the first-order phase transition.

In addition, the resistivity of the low-temperature phase decreases by orders of magnitude when the volume decreases [Fig.~6(a)]. This decrease does not appear to be attributable to an imperfect phase transition: At least in the case of the crystals with $V=3.4\times$10$^6$ and 1.9$\times$10$^4$ $\mu$m$^3$, almost the entire sample volume appears to enter the charge-ordered phase at low cooling rates because the low-temperature resistivity at 0.11 and 0.45 K/min are in good agreement with each other (see Supplementary Fig.~S8). Thus, the decrease in resistivity cannot be straightforwardly accounted for from a transition-kinetics perspective; presumably, another scenario must be employed. For example, the electronic states may vary between the bulk and the surface, which would result in the surface-to-volume-ratio-dependent measured resistivity. This issue is likely related to the microscopic details of the material and is therefore beyond the scope of the present study.

%***********************************************************************
\section{Discussion}

In this study, we have shown that the sample miniaturization in terms of both $V$ and $S$ affects the nucleation kinetics and thus results in a greater degree of supercooling. This size effect may often be involved in a first-order phase transition in thin films and devices because the volume of such systems is substantially smaller than that of bulk crystals. However, when considering $T_{\rm c}^{\rm cool}$ in such systems, one has to note that $T_{\rm c}^{\rm cool}$ is also affected by the thermodynamic $T_{\rm c}$. For example, when a thin film is clamped by the substrate, the lattice constants may differ from those of the corresponding strain-free bulk sample. Moreover, the chemical composition may be unintentionally changed from that of the bulk during fabrication of the thin film or device. These effects can potentially change the thermodynamic $T_{\rm c}$ and thus $T_{\rm c}^{\rm cool}$. Therefore, we refer to them as thermodynamic effects, and they should be considered separately from the kinetic effects that arise from the sample miniaturization and/or temperature-sweeping rates. Provided that $T_{\rm c}^{\rm heat}$ in quantum materials is generally only weakly affected by the kinetic effect (which is at least the case for IrTe$_2$ and $\theta$-RbZn in the present range), $T_{\rm c}^{\rm heat}$ may serve as a facile criterion to determine whether certain thermodynamic effects are involved in $T_{\rm c}^{\rm cool}$ of a small-sized sample of interest.

The size effects may also provide a unique perspective on the relationship between the sample quality and the degree of supercooling. In disordered materials, the low-temperature phase likely consists of multiple domains, the size of which $v_{\rm D}$ is dominated by the sample quality. For the given $i_{\rm v}^{\rm hetero}(T)$ (Fig.~2), the critical cooling rate that separates the slow-cooling and quenched regimes is $\approx$$10V$ for the monodomain situation [Fig.~3(b)], whereas it is $\approx$$10v_{\rm D}$ ($v_{\rm D} \ll V$) for the multidomain situation (Fig.~4). Thus, the low-quality or multi-grain sample is more likely to enter the quenched regime at a given cooling rate than a clean sample with the identical macroscopic volume. In fact, in certain disordered skyrmion-hosting chiral magnets such as Co$_8$Zn$_8$Mn$_4$ \cite{KarubeNatMat} and Fe$_{1-x}$Co$_x$Si \cite{MunzerPRB}, a first-order transition from the magnetic skyrmion lattice (SkL) to conical phases is kinetically avoided even at a low cooling rate of 1 K/min, whereas a cooling rate greater than 2$\times$10$^3$ K/min is required in the nominally clean MnSi with a comparable sample size \cite{OikeNatPhys}. The large difference in critical cooling rate can be attributed partly to the difference in the domain size of SkL. Nevertheless, it should be noted that the disorders generally play multiple roles; depending on the type, the disorders may facilitate heterogeneous nucleation.

%***********************************************************************
\section{Conclusion}

A smaller sample size causes a greater degree of supercooling, and we have experimentally verified this propensity at least qualitatively in two different strongly correlated electron systems: the transition-metal dichalcogenide IrTe$_{2}$ and the organic conductor $\theta$-(BEDT-TTF)$_2$RbZn(SCN)$_4$. When considering the transition behaviors of small-sized first-order-phase-transition systems such as devices and thin films, kinetic effects on supercooling may be involved in addition to thermodynamic effects on $T_{\rm c}$. Our findings also indicate that the critical cooling rate of the material, above which the quenched regime sets in, can be controlled through design of the sample size, potentially facilitating the realization of the quenched state at experimentally accessible cooling rates.

\section*{ACKNOWLEDGMENTS}
	F.K. and H.O. thank T.~Nakajima, T.~Arima, M.~Ikeda, M.~Yoshida, Y.~Okamura, S.~Seki and S.~Shimizu for their valuable discussions. This work was partially supported by JSPS KAKENHI (Grant No. 25220709).


\begin{thebibliography}{99}
\bibitem{Eq1}
The term associated with the surface nucleation may further be divided into the nucleations from the defect-free area and those from defects.
\bibitem{Simulation}
A system that exhibits a quenched metastable state under rapid cooling is supposed to have a peak structure in the temperature dependence of the nucleation rate, as schematically shown in Fig.~1(a). Because the temperature dependence of the nucleation rate in quantum materials is not well established, we employ a standard functional form that has often been used in describing the crystallization process of supercooled liquids. As a model $i_{\rm v}^{\rm hetero}(T)$ for the simulation, we thus consider
\begin{equation} 
i_{\rm v}^{\rm hetero}(T) = a \exp(-\Delta_{\rm k}/k_BT) \exp(-\Delta_{\rm n}(T)/ k_BT),
\end{equation}	
where $\Delta_{\rm k}$ characterizes the temperature-dependent attempt frequency, which is closely related to the diffusion process for classical liquids \cite{Turnbull}, and where $\Delta_{\rm n}(T)$ is the temperature-dependent barrier for the critical nucleation. In the simplest case, $\Delta_{\rm n}(T)$ can be approximated as $\Delta_{\rm n}(T) \propto 1/(T_{\rm c}-T)^2$ (at least near $T_{\rm c}$ \cite{Turnbull}). In the simulations, we set $T_{\rm c} = 100$ K, $\Delta_{\rm k}/k_B$ = 200 K, $\Delta_{\rm n}(T = 0$ K$)/k_B$ = 50 K, and coefficient $a = 1000$.
\bibitem{hosoku} 
In Ref.~\cite{LopezPRB, VO2}, $\int_{T}^{T_{\rm c}}I(T')/rdT'$ in Eq. (3) is phenomenologically approximated as $\int_{T}^{T_{\rm c}}I(T')/rdT' \propto V(T_{\rm c}-T)^n$ to explain the observed size effect in VO$_2$ near $ T_{\rm c}$. This form, however, cannot be extrapolated to sufficiently low temperatures such that the nucleation process, which is innately thermally activated, is strongly inhibited.
\end{thebibliography}
\end{document}